\documentclass[showpacs,preprintnumbers,amsmath,amssymb]{revtex4}

\usepackage{epsfig}
\usepackage{graphicx}
\usepackage{dcolumn}
\usepackage{bm}

\begin{document}

\title{Kernel method for clustering based on optimal target vector}
\author{
Leonardo Angelini, Daniele Marinazzo, Mario Pellicoro, and Sebastiano Stramaglia}
\affiliation{ -TIRES-Center of
Innovative Technologies for Signal Detection and Processing,\\
Universit\`a di Bari, Italy\\
-Dipartimento Interateneo di Fisica, Italy \\
-Istituto Nazionale di Fisica Nucleare, Sezione di Bari, Italy }

\date{\today}

\begin{abstract}
We introduce the notion of {\it optimal target vector}, and describe how it creates a
link between supervised and unsupervised learning. We exploit this notion to construct
Ising models, for dichotomic clustering, whose couplings are (i) both ferro- and
anti-ferromagnetic (ii) depending on the whole data-set and not only on pairs of samples.
The effectiveness of the method is shown in the case of the well known iris data-set and
in benchmarks of gene expression levels.

\end{abstract}

\maketitle

Recent years have been characterized by a dramatic evolution in many fields of life
science with the apparition and rapid spread of so-called high-throughout technologies,
which generate huge amounts of data to handle various aspects of biological samples or
phenomena. The need for efficient methods to represent, analyze and finally make sense
out of these data triggered the development of numerous data analysis algorithms (some of
which physically motivated \cite{domany,angelini}). Among  them, kernel methods have
quickly gained popularity for problems involving the classification and analysis of
high-dimensional or complex data. Well known supervised kernel algorithms are Support
Vector Machines \cite{vapnik}, both for regression and classification, and Kernel Ridge
Regression \cite{shawe}. A popular unsupervised kernel method is Kernel Principal
Components \cite{schol}, a nonlinear projection tool which generalizes the classical
principal component analysis. The purpose of the present work is to propose a simple
connection between supervised and unsupervised kernel methods, which here will permit us
to introduce Ising models for dichotomic clustering.

We recall kernel ridge regression, while referring the reader to \cite{shawe} or
\cite{physiol} for further technical details.  Let us consider a set of $\ell$
independent, identically distributed data $S=\{ ({\bf x}_i, y_i) \}_{i=1}^\ell$, where
${\bf x}_i$ is the $n$-dimensional vector of input variables and $y_i$ is the scalar
output variable. Data are drawn from an unknown probability distribution
$p(\mathbf{x},y)$. The problem of learning consists in providing an estimator
$y=f(\bf{x})$ out of a class of functions, called {\it hypothesis space}. In kernel ridge
regression, calling $\bf{y}$ = $(y_1, y_2, ..., y_\ell)^\top$  the vector formed by the
$\ell$ values of the output variable, the estimator is given by:
\begin{equation}\label{notlinear}
y = f({\bf x}) = \sum_{i=1}^\ell c_i k({\bf x}_i,{\bf x}),
\end{equation}
where coefficients $\{c_i\}$ are given by
\begin{equation}\label{w2}
\bf{c} =  \left (\bf{K} + \lambda \bf{I} \right)^{-1} \bf{y},
\end{equation}
$\bf{K}$ being the $\ell \times \ell$ matrix with elements $K_{ij}=k({\bf x_i},{\bf
x_j})$. $\lambda$ is called regularization parameter, and its main role is to render well
posed the inversion of matrix $\bf{K}$ which in most cases is nearly singular.
$k(\cdot,\cdot)$ is a positive definite symmetric function, and equation
(\ref{notlinear}) may be seen to correspond to a linear estimator in the feature space
$$
\Phi({\bf x}) = ( \sqrt{\alpha_1}\psi_1 ({\bf x}), \sqrt{\alpha_2}\psi_2 ({\bf x}),
...,\sqrt{\alpha_N} \psi_N ({\bf x}), ... ),
$$
where $\alpha_i$ and $\psi_i$  are the eigenvalues and eigenfunctions of the integral
operator with kernel $k$.

Many choices of the kernel function are possible, for example the linear kernel $k({\bf
x},{\bf x'})={\bf x}\cdot{\bf x'}$ leads to the usual linear estimator. The polynomial
kernel of degree $p$ has the form $k({\bf x},{\bf x'})=\left( 1+{\bf x}\cdot{\bf
x'}\right)^p$ (the corresponding features are made of all the powers of ${\bf x}$ up to
the $p$-th). The Gaussian kernel is $k({\bf x},{\bf x'})=\exp{-\left({|| {\bf x}-{\bf
x'}||^2/ 2a^2}\right)}$ and deals with all the degrees of nonlinearity of ${\bf x}$.
Specifying the kernel function $k$ one determines the complexity of the function space
within  which one searches for the estimator, similarly to the effect of specifying the
architecture of a neural network, that is number of layers, number of units for each
layer, etc. Due to (\ref{notlinear}) and (\ref{w2}), the predicted output vector
$\bf{\bar{y}}$, in correspondence of the {\it true} target vector $\bf{y}$, is given by
$\bf{\bar{y}}$$=\mathbf{G}\bf{y}$, where the symmetric matrix $\mathbf{G}$ is given by
\begin{equation}\label{G}
\mathbf{G}=\mathbf{K}\left(\mathbf{K}+\lambda \mathbf{I}\right)^{-1}.\end{equation} Note
that matrix $\mathbf{G}$ depends only on the distribution of $\{\mathbf{x}\}$ values:
$\mathbf{G}$ embodies information about the structures present in $\{\mathbf{x}\}$ data
set. Indeed, the matrix element $G_{ij}$ quantifies how much the target value of the
$j-th$ point influences the estimate of the target of point $i$. The training error $E$
can be calculated as follows $$E
=(\bf{y}-\mathbf{G}\bf{y})^\top(\bf{y}-\mathbf{G}\bf{y})=\bf{y}^\top \mathbf{H}\bf{y},$$
where $\mathbf{H}= \mathbf{I}-2\mathbf{G}+\mathbf{G}\mathbf{G}$ is a symmetric and
positive matrix.

It is worth stressing that, given a kernel function, the corresponding features
$\phi_\gamma ({\bf x})$ are not centered in general. One can show \cite{schol} that
centering the features ($\phi_\gamma \to \phi_\gamma -\langle \phi_\gamma\rangle$, for
all $\gamma$) amounts to perform the following transformation on the kernel matrix:
$$\mathbf{K }\to \mathbf{\tilde{K}}=\mathbf{K}-\mathbf{I}_\ell \mathbf{K}-\mathbf{K}\mathbf{I}_\ell +\mathbf{I}_\ell \mathbf{K} \mathbf{I}_\ell,$$
where $\left( I_\ell\right)_{ij}=1/\ell$, and to work with the centered kernel
$\mathbf{\tilde{K}}$. Therefore in the following we will always assume that the kernel
matrix $\mathbf{K}$ has been centered.

Matrix $\mathbf{G}$ may be seen as a linear convolution filter, mapping target vectors
onto the vector of predicted outputs, and carrying  information about the structures
naturally present in the set of $\{\mathbf{x}\}$ points. In the unsupervised case the
data set is made of $\mathbf{x}$ points, $\{ {\bf x}_i\}_{i=1}^\ell$, and the target
vector $\mathbf{y}$ is missing. However we may consider the following  question: what is
the vector $\mathbf{y}$ such that treating it as the target vector leads to the best fit,
i.e. the minimum training error $\bf{y}^\top \mathbf{H}\bf{y}$? One may expect that this
{\it optimal} target vector would bring information about the structures present in the
data. We look for the optimal target vector in the space of binary functions, ${\bf
\sigma} \in \{-1,1\}^\ell$, the minimizer of the training error $\bf{\sigma}^\top
\mathbf{H}\bf{\sigma}$ thus naturally provides a partition of points in two classes. The
minimizer is the ground state of an Ising model (see, e.g., \cite{kadanoff}) with long
range symmetric couplings $\mathbf{J}$ given by
$$J_{ij}=4G_{ij}-2\sum_{k=1}^\ell G_{ik} G_{kj},$$
for $i\ne j$, and $J_{ii}=0$. We note that, unlike the Potts model introduced in
\cite{domany}, here the couplings are both ferro- and anti-ferromagnetic. Therefore there
is room for frustration and multiple minima. Many algorithms can be used to find an
estimate of the ground state of Ising models, for example simulated annealing \cite{sa}.
Here we use the mean field annealing method \cite{mfa}, which iteratively solves the mean
field equations for the local magnetization vector $\mathbf{m}=\langle
\mathbf{\sigma}\rangle$,
\begin{equation}
\label{mfa} \mathbf{m}=\tanh \left( \beta \mathbf{J}\mathbf{m}\right),
\end{equation}
while decreasing the temperature (increasing $\beta$). The starting temperature
$\beta_{cr}$ may be chosen as the inverse of the maximum eigenvalue of matrix
$\mathbf{J}$. Note that $\mathbf{m} \in \{-1,1\}^\ell$ in the limit of large $\beta$.

Let us now discuss the application to iris data-set, published by Fisher \cite{fisher},
where the sepal length, sepal width, petal length, and petal width are measured in
millimeters on fifty iris specimens from each of three species, {\it iris setosa}, {\it
iris versicolor}, and {\it iris virginica}. Firstly we process all the $150$ points,
using a linear kernel and $\lambda=1$: we find two clusters, one is made of $51$ points
(50 belonging to {\it iris setosa}), the other is made of the remaining $99$ points.
Then, we process the $99$ points with linear kernel and the same value of $\lambda$,
obtaining two clusters of 48 points (47 belonging to {\it iris versicolor}) and 51 points
(49 belonging to {\it iris virginica}). Globally four points are misclassified, with
efficiency of classification $0.973$. In figure 1 we show the four misclassified points.
Results are quite insensitive to $\lambda$: the same classification is obtained varying
$\lambda\in [0.05,100]$. Using a Gaussian kernel we obtain even better efficiency: using
$a=3$ only two points are misclassified, with efficiency $0.987$.

Now we consider application on gene expression data sets. Firstly the colon cancer data
set of \cite{alon}, consisting in 40 tumor and 22 normal colon tissues samples, each
described by 2000 gene expression levels; data are available on the Kent Ridge
Bio-medical Data Set Repository \cite{datab}. We are not going to face, in this place,
the important task of feature selection which is fundamental in the analysis of gene
expression data. The following preprocessing is used to normalize data. First, for each
gene, expression levels are rescaled so as to get unit mean over tissues. Then, for each
tissue, expression levels are linearly transformed to have zero mean (over genes) and
unit variance. Application of our method, using all the genes and a linear kernel
($\lambda =1$), leads to 12 misclassified points with an efficiency of 0.806. Then we
select the $100$ most discriminant genes, using a supervised step where nonparametric
Wilcoxon test is used to asses the capability of each gene in discriminating tumor and
normal tissues. After ranking genes by Wilcoxon test, we apply our method (still with
linear kernel and $\lambda =1$) using only 100 attributes corresponding to the first 100
genes. The output is depicted in figure 2, seven points are misclassified with efficiency
$0.887$. Next we consider the leukemia data set of \cite{golub}, consisting of samples of
tissues of bone marrow samples, $47$ affected by acute myeloid leukemia (AML) and $25$ by
acute lymphoblastic leukemia (ALL), $7129$ attributes. Data are normalized as in the
previous example. Using all genes, and linear kernel, leads to a poor performance
(efficiency 0.569). Ranking genes by Wilcoxon test (assessing the capability of each gene
in discriminating AML and ALL tissues) and selecting the best 500 genes, leads to better
performances: our method with a linear kernel misclassifies 8 points, with efficiency
0.889, see figure 3. Our results are quite insensitive to the choice of $\lambda$.

We now summarize our method. Dichotomic clustering is described as the problem of
evaluating the ground state of Ising models, with couplings values determined in the
frame of kernel methods for supervised learning. The effectiveness of the proposed
approach is shown here on three real examples in terms of high efficiency of
classification. As the output depends very weakly on the regularization parameter
$\lambda$, only the choice of the kernel matters. In this paper we  mostly deal with the
simplest kernel, the linear one: the problem of kernel selection, common to any kernel
method, will be considered elsewhere.

Some remarks are in order. Firstly, it is worth stressing that some information about
structures can also be recovered from diagonal elements $G_{ii}$. In the leave-one-out
scheme \cite{vapnik} a data point $i$ is removed from the data set and the model is
trained using the remaining $\ell -1$ points: let us denote $\tilde{y}_i$ the target
value thus predicted, in correspondence of $\bf{x_i}$. It can be shown that the
leave-one-out error $\tilde{y}_i -y_i$ and the training error $\bar{y}_i -y_i$, obtained
using the whole data set, satisfy:
\begin{equation}
\label{loo} \tilde{y}_i -y_i={\bar{y}_i -y_i \over 1-G_{ii}}.
\end{equation}
This formula shows that the closer $G_{ii}$ to one, the farther the leave-one-out
predicted value from those obtained using also point $i$ in the training stage. If point
$i$ is in a dense region of the feature space then one may expect that removing this
point from the data-set would not change much the estimate since it can be well
"interpolated" by neighboring points. Therefore points in low density regions of the
feature space are characterized by diagonal values $G_{ii}$ close to one, while $G_{ii}$
is close to zero for points $\mathbf{x_i}$ in dense regions. In figure 4 we depict
$\{G_{ii}\}$ for the case of iris.

Secondly, one may look for the optimal target vector $\mathbf{y}$ in $\mathbb{R}^\ell$;
to avoid the trivial solution $\mathbf{y}=\mathbf{0}$, one may constrain $\mathbf{y}$ to
have unit norm, $\mathbf{y}^\top\mathbf{y}=1$. This problem then becomes equivalent to
find the the normalized eigenvector of $\mathbf{H}$ with the smallest eigenvalue. On the
other hand, matrix $\mathbf{H}$ is a function of matrix $\mathbf{K}$: hence it has the
same eigenvectors of $\mathbf{K}$ while the corresponding eigenvalues are related by a
monotonically decreasing function. Therefore, optimal target functions in
$\mathbb{R}^\ell$ coincide with the kernel principal components: this shows that the
method introduced in \cite{schol} may be motivated also as the search for the optimal
target functions.

Finally, as already shown in the iris example, we remark that multiclass clustering may
be obtained by repeated application of the proposed dichotomic method. The notion of
optimal target vector is an interesting bridge between supervised and unsupervised
learning: here we have considered application of this notion for dichotomic clustering.
In our opinion it will carry to the developments of other effective algorithms for the
analysis of complex data.

\begin{figure}[ht!]
\begin{center}
\epsfig{file=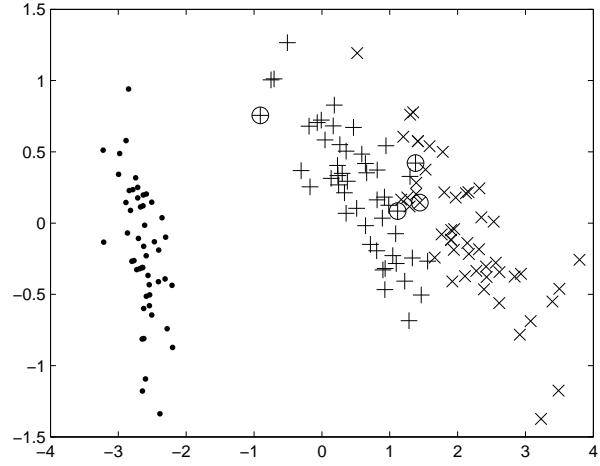,height=7.cm}
\end{center}
\caption{{\small In the plane of the first and second principal components, the iris
data-set is represented. Symbols represent classes: $\cdot$ {\it setosa}, $+$ {\it
versicolor}, $x$ {\it virginica}. The four misclassified points are surrounded by a
circle. \label{fig1}}}
\end{figure}

\begin{figure}[ht!]
\begin{center}
\epsfig{file=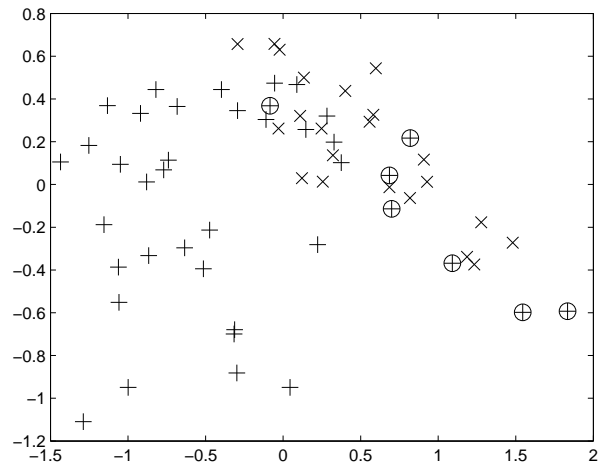,height=7.cm}
\end{center}
\caption{{\small The colon data set, $+$ normal and $\times$ tumor, is represented in the
plane of the first and second principal components (extracted over the $100$ most
discriminating genes). Misclassified points are surrounded by a circle \label{fig2}}}
\end{figure}

\begin{figure}[ht!]
\begin{center}
\epsfig{file=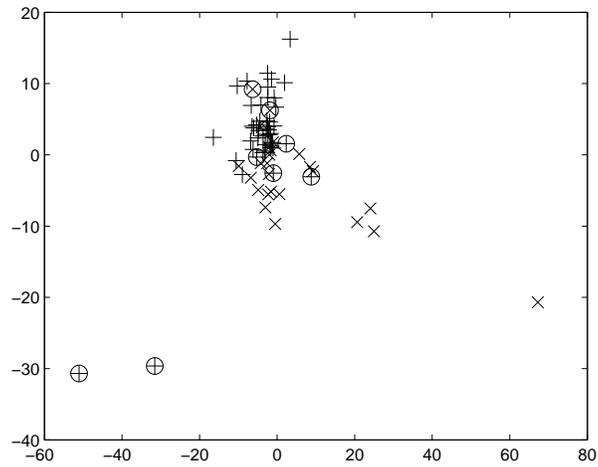,height=7.cm}
\end{center}
\caption{{\small The leukemia data set, $+$ AML and $\times$ ALL, is represented in the
plane of the first and second principal components (extracted over the $500$ most
discriminating genes). Misclassified points are surrounded by a circle. \label{fig3}}}
\end{figure}

\begin{figure}[ht!]
\begin{center}
\epsfig{file=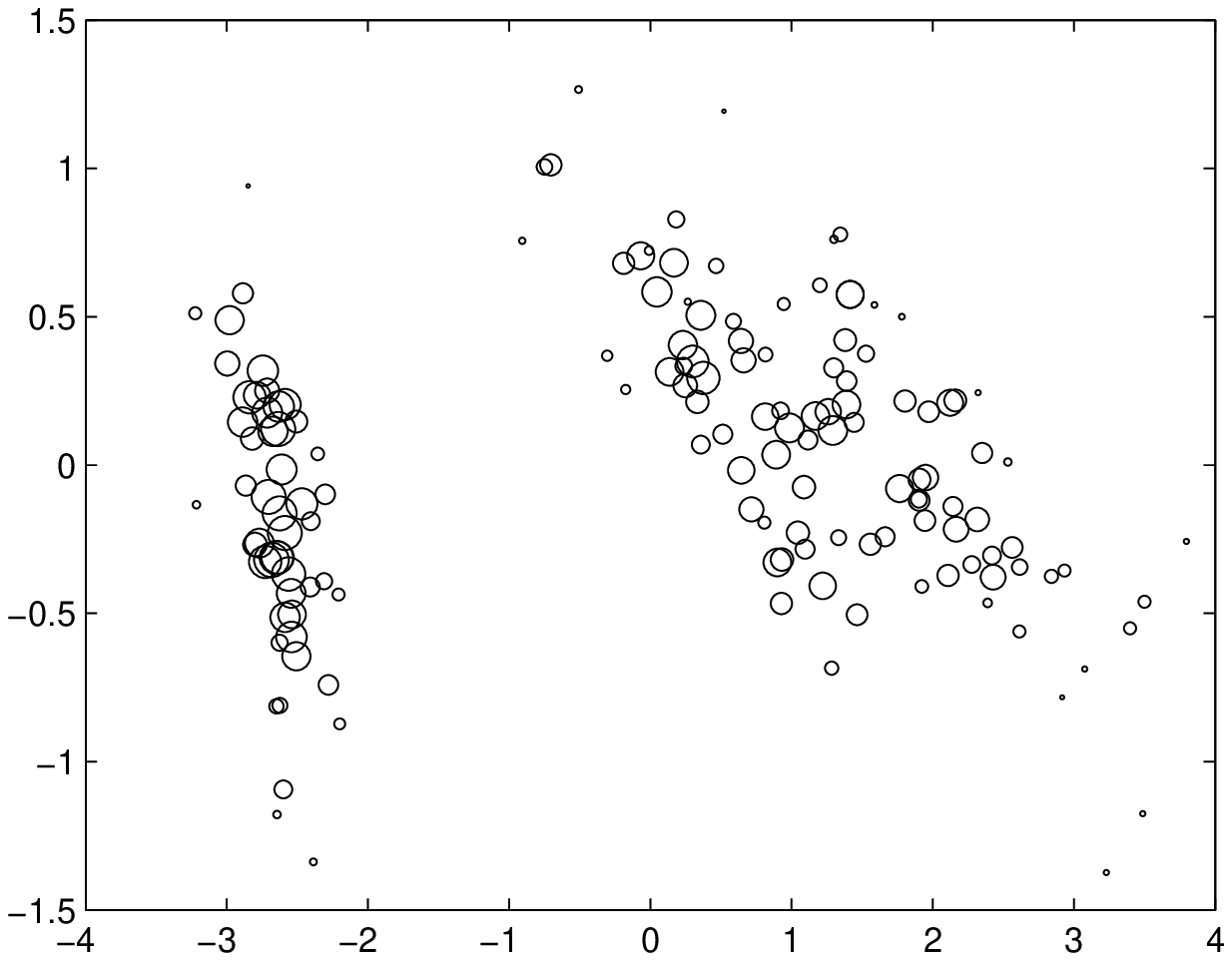,height=7.cm}
\end{center}
\caption{{\small The 150 points of iris data-set are depicted (first and second principal
components). The radius of the circle around each point is proportional to $1-G_{ii}$,
the matrix $\mathbf{G}$ being evaluated using a Gaussian kernel with $\sigma=0.5$ and
$\lambda =1$. \label{fig4}}}
\end{figure}


\begin{thebibliography}{99}
\bibitem{domany} M. Blatt, S. Wiseman, and E. Domany, {\it Phys. Rev. Lett.} {\bf 76}, 3251
(1996).
\bibitem{angelini} L. Angelini, F. De Carlo, C. Marangi, M. Pellicoro, S. Stramaglia, {\it Phys. Rev. Lett.}
{\bf 85} 554 (2000); G. Getz, E. Levine, and E. Domany, {\it PNAS} {\bf 97} 12079 (2000);
L. Angelini, L. Nitti, M. Pellicoro, S. Stramaglia, {\it Phys. Lett.} {\bf A 285} 279
(2001); D. Horn and A. Gottlieb,{\it Phys. Rev. Lett.} {\bf 88} 018702 (2002).
\bibitem{vapnik}V. Vapnik, {\it Statistical Learning
Theory} New York: John Wiley \& Sons, INC (1998).
\bibitem{shawe}J. Shawe-Taylor and N.
Cristianini, {\it Kernel Methods for Pattern Analysis} Cambridge University Press, 2004.
\bibitem{schol}B. Scholkopf, A. Smola, K.-R. Muller, {\it Neural Computation} {\bf 10}
1299 (1998).
\bibitem{physiol}N. Ancona, R. Maestri, D. Marinazzo, L. Nitti, M.
Pellicoro, G.D. Pinna, S. Stramaglia, {\it Physiol. Meas.} {\bf 26} 363 (2005).
\bibitem{kadanoff}L.P. Kadanoff, {\it Statistical Physics: Statics, Dynamics
and Renormalization} World-Scientific, 2000.
\bibitem{sa} S. Kirkpatrick, C. D. Gelatt Jr., M. P. Vecchi, {\it Science} {\bf 220} 671 (1983).
\bibitem{mfa}J.J. Hopfield and D.W. Tank, {\it Biol. Cybern.} {\bf 52} 141 (1985); C. Peterson and B. Soderberg, {\it Int. J. Neural Syst.} {\bf 1} 3 (1989).
\bibitem{fisher}R. A. Fisher, {\it Annals of Eugenics} {\bf  7} 179 (1936).
\bibitem{golub} T.R. Golub et al., {\it Science} {\bf 286} 531 (1999).
\bibitem{datab} http://sdmc.lit.org.sg/GEDatasets/Datasets.html
\bibitem{alon} U. Alon et al., {\it PNAS} {\bf 96} 6745 (1999).


\end{thebibliography}
\end{document}